# ENHANCED THERMAL CONDUCTIVITY IN NANOFLUIDS UNDER THE ACTION OF OSCILLATING FORCE FIELDS

Clément Le Goff, Philippe Ben-Abdallah[*] , Gilberto Domingues and Ahmed Ould El Moctar
Laboratoire de Thermocinétique, CNRS UMR 6607, Ecole Polytechnique de l'Université de Nantes, 44 306 Nantes cedex 03, France ; * Author for correspondence  (Tel. : 33 (0) 2 40 68 31 17 ; Fax : 33 (0) 2 40 68 31 17 ; E-mail: pba@univ-nantes.fr)

**Abstract** The thermal conductivity of nanoparticles colloidal suspensions, submitted to the action of an external force field has been calculated by non equilibrium molecular dynamics simulations. For driven forces in the radio frequency and microwave ranges, we show that the thermal conductivity of nanofluids can be strongly enhanced without cluster formation.

**Key words**  Non equilibrium molecular dynamics. Active control of transport coefficients. Nanocolloïds. Cooling system

## Introduction

Nanofluids (Eastman et al. 2001, 2004) are liquids containing suspended nanometer-size solid particles. Many of these colloidal suspensions have displayed (Eastman et al. 2001; Lee et al. 1999; Das et al. 2003) in many circumstances an anomalously high thermal conductivity that the classical theory of effective media (Hamilton and Crosser 1962) fails to explain without clusters in the suspension. This unexpected result has caused a strong interest in the scientific community (see Keblinski  et al. 2005 and and Prasher et al. 2005 for a recent review) because it paves the way to a wide variety of potential applications ranging from the transportation and energy production to promising fluidic cooling technologies. Considerable efforts have been developped, since then (see for example Vladkov and Barrat 2006; Ben-Abdallah 2006;  Eapen  et al. 2007; Prasher et al. 2005; Prasher et al. 2006; Evans et al. 2006) to understand the origin of this effect. However, today, this problem still remains open.

Besides, here we question on the possibility of enhancing the heat transport coefficient of a nanofluid from external actuations. It is now well known that nanoparticles dispersed in a solvent can easily be manipulated by external forces which bring them into nonequilibrium situations in a controllable manner (Ashkin et al. 1986 and Refs. therein). One of the simplest ways to  generate such a force field is, for example, done by  focusing a laser beam with an appropriate wavelength on the colloidal suspension which provides a momentum to the nanoparticles. In this work we study in what extent the motion of nanoparticles, submit to the action of an external oscillating force field, affect the thermal conductivity of nanofluids. Two different situations are investigated in this work. In the first one, a temperature gradient is established through the solvent and an oscillating force field is applied in the same direction. Then, using the non equilibrium molecular dynamics technique (Koplik and Banavar 1995), we calculate the thermal conductivity of fluid in the direction of force field and in the transversal direction. For periodic excitations in the radio frequency and microwave ranges, we show that the thermal conductivity of nanofluid can be enhanced. We identify the best frequency conditions to enhance heat transport in nanofluids and we examine whether the thermal conductivity of nanofluids can be carried out without clustering in presence of external forces.

## Monodisperse two-dimensional nanofluid model

In the present study we perform our simulations on two dimensional nanofluids (fig.1) which consists on N identical crystalline particles of mean radius R dispersed in a Lennard-Jones fluid of density $\rho = N/S = 0.7$, N being the number fluid particles (20000) and S the surface of the simulation box. Two types of potentials are used to describe the interactions between fluid atoms, solid atoms and between the fluid and solid phase. The first one is a truncated Lennard-Jones (TLJ) potential which acts between all pairs of fluids particles as,

$$U_{TLJ}(r) = \begin{cases} U_{LJ}(r) - U_{LJ}(r_c) & r \leq r_c \\ 0 & r > r_c \end{cases}, \qquad (1)$$

where $U_{LJ}(r) = 4\varepsilon[(\frac{\sigma}{r})^{12} - (\frac{\sigma}{r})^6]$ is a 12-6 LJ potential, $r = |r_i - r_j|$ stands for the interparticle distance while $\varepsilon$, $\sigma$ and $r_c = 1.5\sigma$ represent the LJ parameters and the cutoff radius of TLJ potential, respectively. The cohesion forces between solid atoms are ensured by the Finite extension non-linear elastic (Fene) potential,

$$U_{Fene}(r) = \frac{k}{2} R_0^2 \ln[1 - (\frac{r}{R_0})^2]. \qquad (2)$$

Here, $k = 17.8\varepsilon/\sigma^2$ is the strength factor and $R_0 = 1.5\sigma$ is the maximum allowed length of an atomic bond. The solid particles are constructed by carving solid disk in an hexagonal bulk crystal. Each particle contains $N_s=151$ atoms and have mean radius $R \approx 6\sigma$. The nanoparticle volume fraction is equal to 4% which corresponds, for a box of length $L = 170\sigma$, to a total number of 16 nanoparticles. The interactions between the atoms of solid particles and the atoms of fluid are also described by a LJ potential analog to (1) but with a cutoff radius $r_c = 2.5\sigma$.

In this paper we assume that the nanoparticles are able to couple with an external oscillating field throughout a force

$$F_{ext} = F_0 \cos(2\pi f \times t). \qquad (3)$$

which acts on each particles atom. To estimate the magnitude $F_0$ of this force we need to overcome the drag force exerted on each nanoparticle by the fluid particles collision while respecting the applicability of periodic boundary conditions, we have performed some preliminary calculations. First, the drag force has been estimated applying the Stokes Einstein theory. In this framework, the diffusion coefficient of particles reads in terms of the autocorrelation function of the nanoparticle velocity (see fig.2) as

$$D = \frac{1}{2} \int_0^\infty \langle v_{NP}(t).v_{NP}(0) \rangle dt, \qquad (4)$$

where $v_{NP}$ denotes the velocity of the mass center of the nanoparticle. Then, according to the Stokes Einstein theory, the viscosity of two-dimensional fluid writes

$$\eta = \frac{k_B T}{3\pi D}. \qquad (5)$$

Hence, the drag force exerted on a nanoparticle in motion with a velocity $v_{NP}$ is given by

$$F_{drag} = 3\pi \eta v_{NP}. \qquad (6)$$

For the nanoparticles we have previously introduced we have found a diffusion coefficient $D = 0.06$, a viscosity $\eta = 1.75$ and a drag force equal to $F_{drag} = 1.1$ [the mean velocity we have used has been obtained by MD and is equal to $v_{NP} = 6.48 \times 10^{-2}$] in LJ units with $\varepsilon = 1.6 \times 10^{-21} J$ and $\sigma = 3.4 \times 10^{-10} m$. In parallel, in order to respect the applicability of periodic boundary conditions, the magnitude $F_0$ of the net force applied on each particle have been chosen so that their displacement over a period of this oscillating force is smaller than the length of the half simulation box. But, it is well known, in a mechanical point of view, that the dynamics of a single (isolated) particle in presence of an external force field in a fluid is governed by the Langevin equation $m_{NP}\ddot{y} = -\gamma\dot{y} + N_s F_0 \cos(\omega t) + F_B$, where $\gamma = 3\pi\eta$ represents, in the Stokes approximation, the friction factor due to the fluid while $F_B(t)$ is the Brownian force which can be regarded as a Gaussian white noise. In the frequency range under consideration in this study (the radio frequency range), $f_{min} = 1/300\pi$ and $f_{max} = 2/3\pi$ in LJ unit, one easily verified that, the driving force must obey to the following relation

$$N_s F_0 < F_{max} \approx 9.6 \qquad (7)$$

to respect this condition. In our simulation, to simultaneously respect the conditions $N_s F_0 > F_{drag}$ and $N_s F_0 < F_{max}$, we have set $F_0$ to $5.1\times 10^{-2}$ which corresponds to a global force $N_s \times F_0$ acting on each nanoparticle equal to 7.7 (LJ unit). MD simulations were performed with the Verlet algorithm using periodic boundary conditions in the two dimensions of space. For our numerical experiments the integration time step $\Delta t$ is set to 0.005 LJ time and the typical lengths of a run is equal to $1.5\times 10^6$ MD time steps. During the $10^5$ first time steps the system is equilibrated at a temperature $T=1$ (LJ unit), using the following velocity rescaling in each phase,

$$v_{new} = [\frac{N_p}{\sum_i m_i v_i^2}]^{1/2} v_{old} , \qquad (8)$$

where $N_p$ is the number of particles in the phase p.

**Heat transport coefficient and non-equilibrium molecular dynamic simulations**

After this equilibration step we start the non equilibrium simulation by creating a temperature gradient in a direction parallel to one of the sides of the simulation box. To do that we first define three distinct regions in the box (fig. 1). The two first ones are used as heat sources and dissipate with the same power $P = \frac{dQ}{dt} = 6$ by unit of LJ time. They are located on two opposed faces of the box over a region of thickness 3 (LJ unit). The second region is a sink 3 wide (LJ unit) which is located in the center of the simulation box. In this region, heat leaves the system with a global power $2P$ so that the same quantity of energy is added to the system and subtracted from it. After having discretized the simulation box into thin layers 4 wide (LJ unit) we apply and cut out heat flux as described above during $2\times 10^5$ MD time steps. After this preliminary step, we apply the oscillating force on the nanoparticles cloud. Upon the application of this force, we let the system to acquire a steady state during a $10^5$ time steps. Then, we record the temperature profile during about $1.2\times 10^6$ MD time steps to finally calculate the time average of the temperature field (fig.3) in the simulation box. To avoid an overheating of nanofluid due to the power dissipation from the external force we also apply a velocity rescaling each $10^3$ MD time steps. This allow us to work, at any position, at constant temperature without perturbing the temperature gradient.

By definition a nanofluid is a binary mixture liquid-solid. In the local thermodynamic equilibrium assumption the mass flux $\mathbf{J}_m$ related to the diffusion of fluid (resp. solid particles) and the heat flux $\mathbf{J}_q$ are related to the local affinities $\mathbf{X}_m = -\frac{1}{T}\nabla\mu$ ($\mu$ is the chemical potential) and $\mathbf{X}_q = -\frac{1}{T^2}\nabla T$ by the linear relations (Barrat and Hansen 2003) :

$$\mathbf{J}_m = L_{mm}\mathbf{X}_m + L_{mq}\mathbf{X}_q, \tag{9}$$

$$\mathbf{J}_q = L_{qm}\mathbf{X}_m + L_{qq}\mathbf{X}_q. \tag{10}$$

where $L_{ij}$ are the Onsager coefficients. If we substitute the thermodynamic force $\mathbf{X}_m$ from (9) into relation (10) then, according to the Onsager's reciprocity relations (i.e. $L_{mq} = L_{qm}$), the heat flux writes

$$\mathbf{J}_q = \frac{L_{qm}}{L_{mm}}\mathbf{J}_m + [L_{qq} - \frac{L_{qm}^2}{L_{mm}}]\mathbf{X}_q. \tag{11}$$

In the absence of diffusion (i.e. $\mathbf{J}_m = 0$) the heat current is given by the famous Fourier's laws

$$\mathbf{J}_q = -\lambda \nabla T \tag{12}$$

where $\lambda = \frac{1}{T^2}\left(L_{qq} - \frac{L_{qm}^2}{L_{mm}}\right)$. In that case, a combined measure of heat flux and of temperature gradient allow us to obtain the thermal conductivity of nanofluid. In our numerical experiments we have verified that the diffusion of solid particle could be neglected in presence of sufficiently intense oscillating force field. Thus, after having collected the average temperature profile we have calculated (fig.3), in linear regime (low dissipated power), the thermal conductivity of nanofluid using the Fourier's law

$$J_q = -\lambda \frac{\partial \langle T \rangle}{\partial y} = -\lambda \frac{T_{\sin k} - T_{source}}{L/2}, \tag{13}$$

where $J_q = P/L$ is the linear density of flux at y=0. To provide a basis for comparison we have also performed calculations of the thermal conductivity of fluid alone and of nanofluid without external excitation. We have found, for the former, a thermal conductivity equal to $6.59$ (LJ unit) while it is equal to $6.52$ (LJ unit) for the second.

**Results and discussion**

Now, let us turn to the discussion of our numerical results. Two sets of experiments have been carried out in this work. In the first one, we study the thermal conductivity $\lambda_{//}$ of nanofluid in the direction parallel to an external periodic force field versus its oscillating frequency. The results display on fig. 4 show, in a counterintuitive way, that the thermal conductivity does not evolves monotonically with this frequency. Indeed, we clearly distinguish two singularly different thermal regimes depending on the frequency of the force field. Below a low frequency limit, approximately equal to $2 \times 10^{-3}$ (LJ unit) we see that the thermal conductivity drastically increases. In this "low frequency regime", a collective transport of fluid particles driven by the oscillating force has been highlighted (Fig. 5). On this figure we have plotted some characteristic trajectories of fluid particles in this regime. It clearly appears that the trajectories of fluid particles is very similar to that of nanoparticles . On the contrary, at high frequency, the hydrodynamic behavior of fluid radically changes. In fact, for frequencies outside of the low regime, the hydrodynamics effects tend to disappear and, consequently, we observe a weaker enhancement of the thermal conductivity of nanofluid. Moreover, as the frequency of the force increases to tends to the microwave range we see that the thermal conductivity decreases but does not tends to its value "at rest" that is without excitation. At the present time, we do not completely understand this result. It could be explain by a modification of the thermal conductivity of nanoparticles themselves due to a resonant excitation of acoustic modes of atoms. However, this point needs further investigations to be clarified.

The second numerical study performed concerns the evaluation of heat transport properties of nanofluid in the transversal direction of the driven force. Here again, we observe the same thermal regimes as described above. However, the thermal conductivity enhancement is much less important than in the parallel configuration. This result can easily be interpreted as follow. When a nanoparticle moves in a direction perpendicular to the temperature gradient the fluid dragged by this particle less participates to the heat exchange between the hot source and the sink.

Another interesting result we have highlighted in all simulations we have performed with an oscillating force in the radio frequency and microwave ranges is the absence of cluster formation. This result is radically different to what we have already observed (Ben-Abdallah 2006) in presence of steady force fields and allow to avoid the negative effects of clogging in nanofluidic devices. With an oscillating force field, we have observed that during an half period of the force many nanoparticles tend to stick together. However, we have noticed that these particles separate as soon as the sign of the force changes. Here, we do not discuss furthermore this complex mechanism and let this discussion to future works.

**Conclusions**

In this work, we have demonstrated that the heat transport coefficient of nanofluids can be enhanced by external actuations. This enhancement has been closely related to the hydrodynamic behavior of fluid. When the fluid is collectively driven by the motion of nanoparticles we have shown that the thermal conductivity of nanofluid can be strongly enhanced both in the direction of force field and in the transversal direction. We hope this work will contribute to the development of a new generation of active cooling devices.

# Figure captions

Fig. 1 : Snapshot of atomic positions of the nanofluid model in a non equilibrium situation. A heat flux +Q/2 is added in the sources regions and removed in the central sink .

Fig. 2 : Normalized velocity autocorrelation function of nanoparticles inside a LJ fluid without external force field (solid curve) and in presence of external oscillating force field (dashed curve) in the microwave range ($1.47 \times 10^9 Hz$).

Fig. 3 : Temperature profile : (a) in a LJ fluid (circle) of density $\rho = 0.7$ under the action of a weak heat flux + Q crossing the upper and the lower surfaces $y = \pm 0.5$ in direction of fluid and of a negative heat flux –Q at $y = 0$. (b) in a nanofluid (triangle) with nanoparticles 2nm radius dispersed in a LJ fluid (4% vol.) under the action of an oscillating force field with a frequency in the radio frequency range ($5.8 \times 10^8 Hz$). Temperatures and positions are given in LJ units.

Fig. 4 : Thermal conductivity of nanofluid made with nanoparticle 2nm dispersed (4% vol.) in a LJ fluid in the direction parallel and perpendicular to the excitation. The dashed horizontal line stands for the thermal conductivity of nanofluid without excitation.

Fig. 5 : Characteristic trajectory of fluid particles versus time in low frequency and high frequency regime. The period of oscillating excitation is $T_{lf} = 785$ in low frequency regime and $T_{hf} = 314$ in high frequency regime. Positions are given in LJ units. The oscillating force field is plotted for comparison in dashed line. Positions, force magnitude and time are given in LJ units.

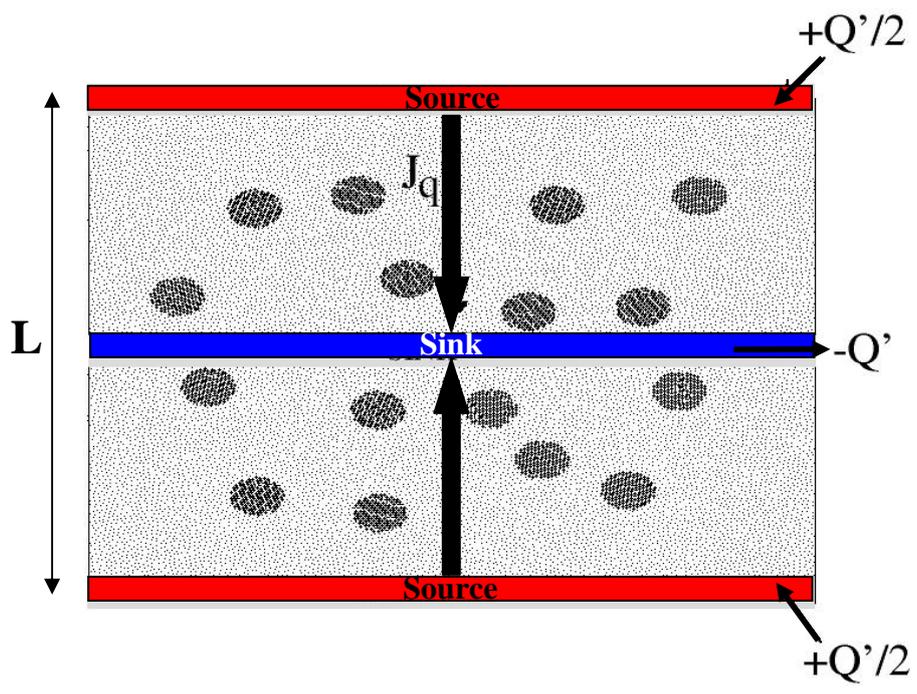

**Figure 1**

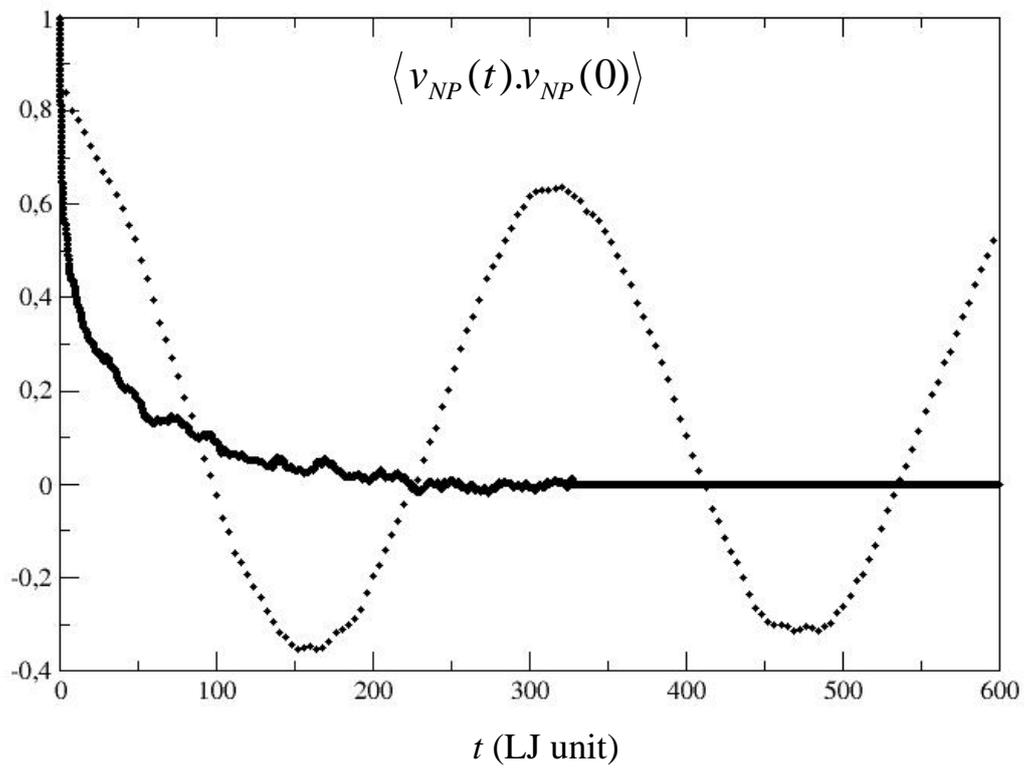

**Figure 2**

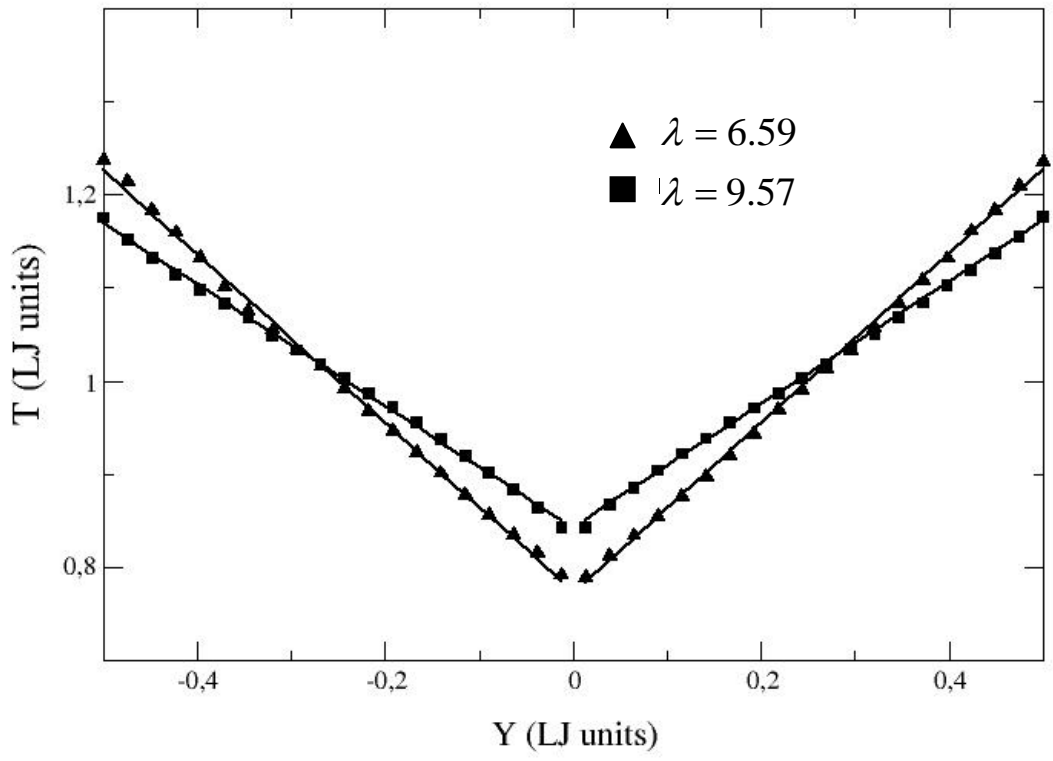

**Figure 3**

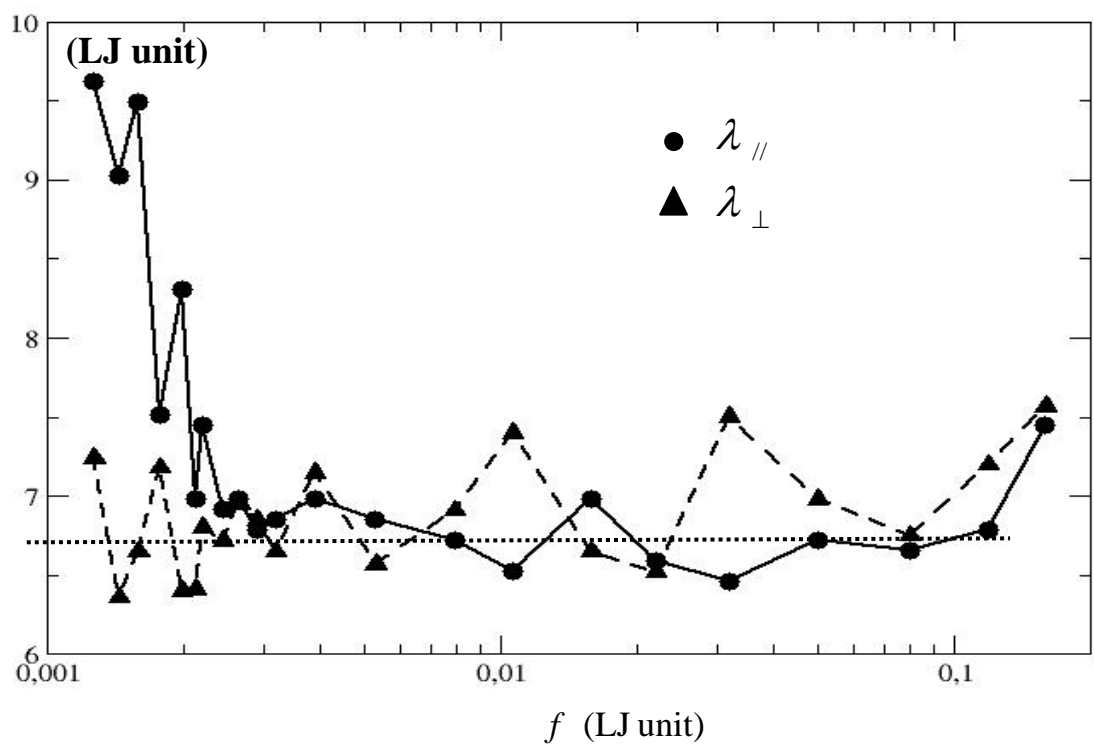

**Figure 4**

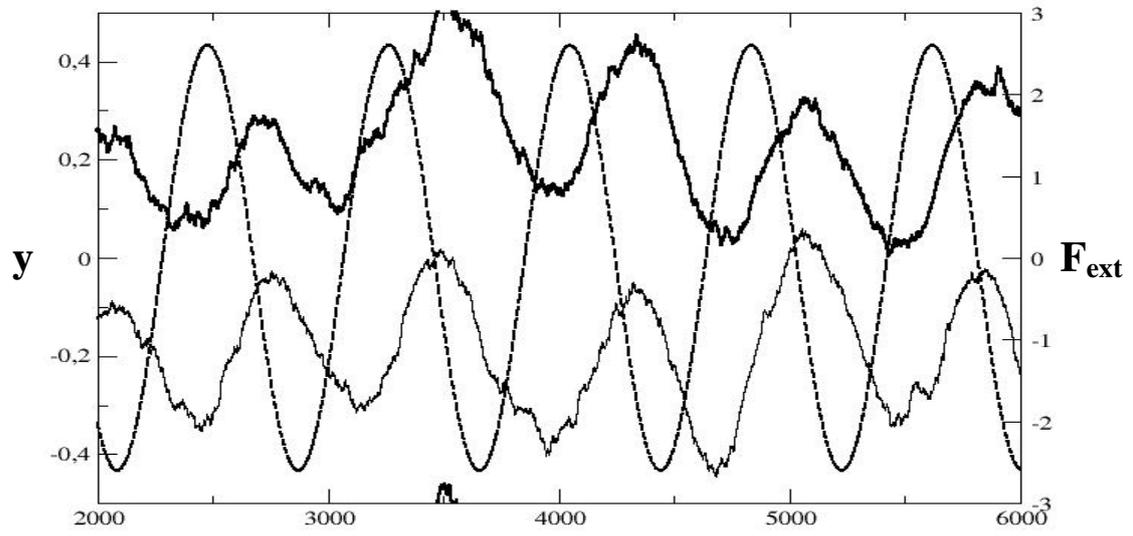

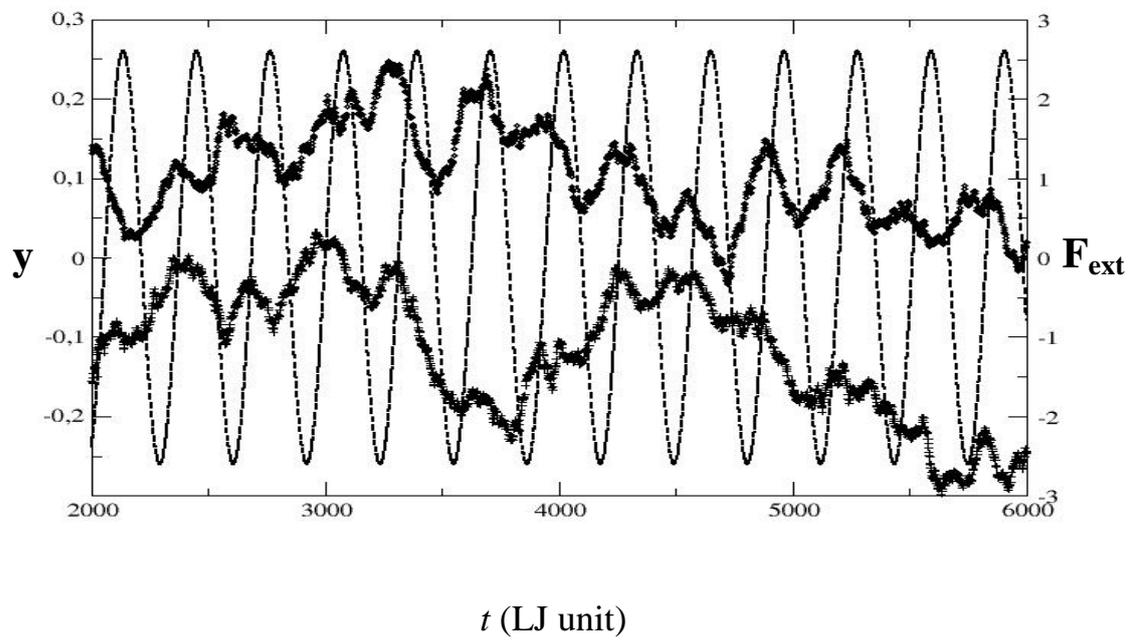

*t* (LJ unit)

**Figure 5**